\input{aipcheck}

\documentclass[
    ,final            % use final for the camera ready runs
  ]
  {aipproc}

\layoutstyle{8x11double}
\usepackage{bm}
\usepackage{amssymb}

\begin{document}

\title{\vspace{-2cm}
\rightline{\normalsize\rm ANL-HEP-CP-10-6}
\vspace{1.3cm}
Closing a Loophole in Factorization Proofs}


\author{Geoffrey 
T.~Bodwin\footnote{Speaker.}$^{~,}$}
{
address={HEP Division, Argonne National Laboratory, 
9700 South Cass Avenue, Argonne, Illinois, 60439, USA}
}

\author{Xavier Garcia i Tormo}{
  address={HEP Division, Argonne National Laboratory,
9700 South Cass Avenue, Argonne, Illinois, 60439, USA}
,altaddress={
Department of Physics, University of Alberta, Edmonton, 
Alberta, Canada, T6G 2G7~\footnote{Current address.}}
}

\author{Jungil Lee}{
address={Department of Physics, Korea University, Seoul 136-701, 
Korea}
%,altaddress={ and Korea Institute of Science and Technology Information, 
%Daejeon 305-806, Korea}
}

\begin{abstract}
We address the possibility in factorization proofs that low-energy
collinear gluons can couple to soft gluons.
\end{abstract}

\maketitle

%%%%%%%%%%%%%%%%%%%%%%%%%%%%%%%%%%%%%%%%%%%%
%% MAINMATTER
%%%%%%%%%%%%%%%%%%%%%%%%%%%%%%%%%%%%%%%%%%%%

This talk is based on Ref.~\cite{Bodwin:2009cb}, in which one can find a
more detailed discussion.

The goal in factorization of QCD processes is to separate perturbative
processes at the scale of the large momentum transfer $Q$ from
nonperturbative processes at the scale of $\Lambda_{\rm QCD}$ or smaller.
In a factorization formula, the perturbative contributions are contained
in short-distance coefficients, which are process dependent. The
nonperturbative contributions are contained in long-distance quantities,
such as parton distribution functions and fragmentation functions. The
predictive power of factorization formulas comes from the process
independence (universality) of the nonperturbative quantities.

For hard-scattering processes in QCD, the nonperturbative contributions
arise from the emission of soft gluons or gluons that are collinear to
external particles.  These gluons and the associated propagators to
which they attach have virtualities that are much less than the
hard-scattering scale $Q$. We use the light-cone momentum components 
$k=(k^+,k^-,\bm{k}_\perp)$, with $k^{\pm}=(1/\sqrt{2})(k^0\pm k^3)$.
Then the components of the momentum of a soft gluon have the 
orders of magnitude
\begin{equation}
k_S \sim Q\epsilon_S(1,1,\bm{1}_\perp),
\end{equation}
where $\epsilon_S \ll 1$. There is a soft logarithmic singularity that
is associated with the limit $\epsilon_S \to 0$.  Suppose that the
momenta of the external particles are along the $\pm$ light-cone
directions. Gluons that are collinear to these external particles have
collinear-to-plus ($C^+$) and collinear-to-minus ($C^-$) momenta:
\begin{eqnarray}                                                     
k_{C^+} &\sim& Q\epsilon^+[1,(\eta^+)^2,\bm{\eta}^+_\perp],\\
k_{C^-} &\sim& Q\epsilon^-[(\eta^-)^2,1,\bm{\eta}^-_\perp],
\end{eqnarray}                                                       
where $\eta^\pm \ll 1$. There is a collinear logarithmic singularity
that is associated with the limit $\eta^\pm \to 0$. There is also a soft
logarithmic singularity that is associated with the limit $\epsilon^\pm
\to 0$.

\begin{figure}
  
\includegraphics[width=0.5\columnwidth]{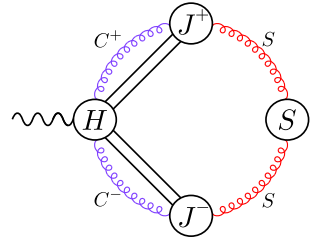}
  \caption{Conventional leading regions for $e^+e^-$ annihilation into 
   two light mesons.}
  \label{fig:conv-leading}
\end{figure}

Leading regions are the Feynman-diagram topologies that yield
contributions that are leading in powers of $Q$. One can find the
leading regions for gauge theories by analyzing pinch singularities in
the momentum contours of integration and by making use of power-counting
arguments
\cite{Sterman:1978bi,Sterman:1978bj,Libby:1978qf,Collins:1985ue,Collins:1989gx}.
For definiteness, consider $e^+e^-$ annihilation into two light mesons.
In this discussion and in subsequent discussions, we work in the Feynman
gauge. The conventional leading regions have the form that is shown in
Fig.~\ref{fig:conv-leading}.
$J^\pm$ are jet subdiagrams, which contain the external particles and
associated collinear gluons.  $S$ is a soft subdiagram, which contains
soft gluons.  $H$ is a hard subdiagram, which contains only propagators
with virtuality of order $Q^2$.  In the conventional picture of the
leading regions, soft gluons attach to the collinear subdiagrams and
collinear gluons attach only to the hard subdiagram. This form is also
implicit in the action in soft-collinear effective theory (SCET) 
\cite{Bauer:2000yr}.

\begin{figure}
  \includegraphics[width=1.0\columnwidth]{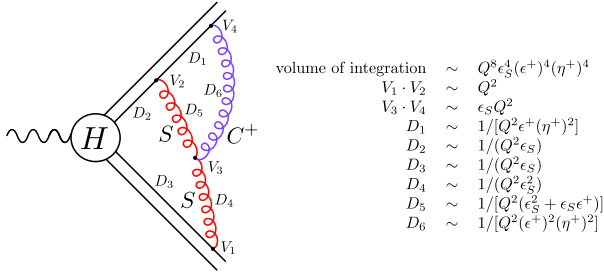}
  \caption{Two-loop example of a $C^+$ gluon attaching
 to a soft gluon.}
 \label{fig:two-loop-ex}
\end{figure}

\begin{figure}
  \includegraphics[width=0.5\columnwidth]{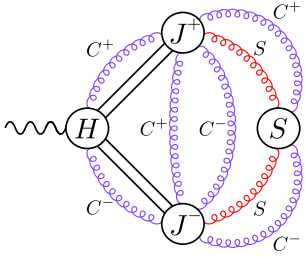}
  \caption{Leading regions for $e^+e^-$ annihilation into 
   two light mesons.}
  \label{fig:leading-regions}
\end{figure}

However, low-energy collinear gluons can couple to soft gluons. Consider
the two-loop example in Fig.~\ref{fig:two-loop-ex} in which a $C^+$
gluon attaches to a soft gluon.
The vertex, propagator and phase-space factors give (for
$\epsilon^+\lesssim \epsilon_S$) the factor
$\epsilon_S\epsilon^+/(\epsilon_S^2+\epsilon_S\epsilon^+)$. This factor
is independent of $Q$ and  gives a leading contribution if $\epsilon^+
\sim \epsilon_S$. Hence, the leading regions must include couplings of
collinear gluons to soft gluons. Power-counting arguments also show that
low-energy collinear gluons can couple to each other, as well as to the
hard subdiagram. Thus, the leading regions have the form that is shown in
Fig.~\ref{fig:leading-regions}.

Because of color confinement, gluons with momentum components less than
of order $\Lambda_{\rm QCD}$ are unphysical. Therefore, one might ask why we
need to consider gluons with energy less than $\Lambda_{\rm QCD}$.
However, as we have seen, low-energy gluons can appear in perturbation
theory in leading power in $Q$. Perturbative calculations are used to
compute short-distance coefficients. In order to establish the
consistency of such calculations, it is necessary to prove that the
contributions from low-energy gluons can be re-organized into the
standard factorized form. Specifically, for $e^+e^-$ annihilation into two light
mesons, we need to show that the
amplitude factors into (1) a hard function that contains only propagators
with virtualities of order $Q^2$, (2) jet functions that contain all of
the collinear contributions, and (3)  a soft function that contains all
of the soft contributions and that cancels when one sums over
connections to quark and antiquark in a meson. Our strategy for proving
this factorization has the following steps: (1) we show that the soft
and collinear singularities decouple from the hard subdiagram and from
each other; (2) we show that the soft singularities cancel; (3) in the
jet functions, we extend the ranges of integration of the collinear
gluon momenta to regions of order $Q$ around the collinear singularities,
thereby incorporating all of the collinear contributions into the jet
functions; (4), we re-define the hard function to be the amplitude
divided by the extended jet functions. The re-defined hard function is
free of soft and collinear singularities and depends only on the scale
$Q$. Hence, it contains only virtualities of order $Q^2$.

In analyzing the soft and collinear singularities, we need to consider
the possibility that different loop momenta can approach the soft and
collinear limits at different rates.  The allowed limiting procedures
are governed by power-counting arguments. Along a given line, the
momentum components of gluons that attach to the exterior provide lower
bounds on the momentum components of gluons that attach to the interior.
That is, the exterior divergences ``control'' the interior divergences.

We now describe the technical tools that we need in order to prove
factorization: collinear approximations, the soft approximation, and
decoupling relations.

If a gluon carrying $C^\pm$ momentum $k$ attaches to a line that does
not carry $C^\pm$ momentum, then the $C^\pm$ approximation can be
applied \cite{Collins:1985ue,Bodwin:1984hc}. In the $C^\pm$
approximation one replaces the $g_{\mu\nu}$ in the gluon-propagator
numerator with $k_\mu n_{\nu}^\mp/k\cdot n^\mp$, where $n^\mp$ is a
light-like vector in the $\mp$ direction. In the $C^\pm$ approximation,
the index $\mu$ attaches to the non-$C^\pm$ line. The collinear
approximations are exact at the collinear singularities $\eta^\pm=0$.
The collinear approximations do not depend on the momentum of the line
to which the collinear gluon attaches. As we shall see, a very useful
property of the collinear approximations is that they result in a
longitudinal gluon polarization.

%\subsection{Soft Approximation}

If a gluon carrying a soft momentum $k$ attaches to a line carrying
momentum $p$ and the components of $k$ are much less than the largest
component of $p$, then the soft approximation applies
\cite{Grammer:1973db,Collins:1981uk}. The soft approximation consists of
replacing the $g_{\mu\nu}$ in the soft-gluon propagator numerator with
$k_\mu p_\nu/k\cdot p$. The index $\mu$ attaches to the line with
momentum $p$. The soft approximation is exact at the soft singularity
$\epsilon_S=0$. In contrast with the collinear approximation, the soft
approximation depends on the momentum of the line to which the singular
gluon attaches. The soft approximation also results in a longitudinal
gluon polarization.

%\subsection{Decoupling Relations}

\begin{figure}
  \includegraphics[width=1.0\columnwidth]{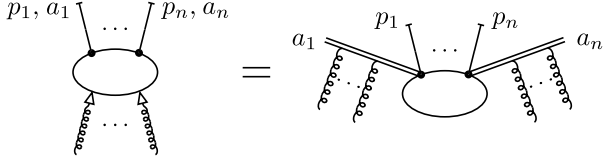}
  \caption{Decoupling relations.}
  \label{fig:decoupling}
\end{figure}

If the longitudinally polarized gluons that result from one of the above
approximations (soft, $C^+$, or $C^-$) attach in all possible ways to a
subdiagram, then the graphical Ward-Takahashi identities can be used to
show that they decouple
\cite{Collins:1985ue,Bodwin:1984hc,Collins:1981uk}. The decoupling
relations are depicted in Fig.~\ref{fig:decoupling}
The hash-marks on the external lines indicate that those lines are
truncated. There can also be an arbitrary number of on-shell external
lines, which are not shown explicitly. The arrows represent the
gluon-propagator replacement factors from the soft, $C^+$, or $C^-$
approximation.  The ``eikonal'' (double) lines  have vertices of the form
$n_\mu$ and propagators of the form $1/(k\cdot n)$, where $n$ is the
vector that appears in the soft, $C^+$, or $C^-$ approximation. These
eikonal lines are path-ordered exponentials of path integrals of gauge
fields.

In non-Abelian gauge theories, the decoupling relations require that the
gluons have momenta that are proportional to each other. This is
automatically the case for gluons with momenta at a $C^+$ or $C^-$
singularity. If a soft gluon with momentum $k$ attaches to a subdiagram
in which all lines have $C^\pm$ singular momenta, then only the
component $k^\mp$ enters into the interactions in the subdiagram.
Without loss of accuracy, we can replace $k$ in the subdiagram and in
the soft approximation with a vector whose only nonzero component is
$k^\mp$ . Then, all of the soft gluons that couple to the $C^\pm$
singular subdiagram have momenta that are proportional to each other, as
is required by the decoupling relation.

\begin{figure}
  \includegraphics[width=0.5\columnwidth]{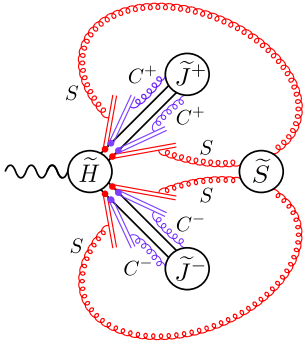}
  \caption{Factored soft and collinear singular contributions.}
  \label{fig:factored-form}
\end{figure}

In order to carry out the factorization, we follow an iterative
procedure, starting with the singular contributions that are innermost
in the Feynman diagrams (those with the largest energy scale) and
working to the outside. Each stage in the iteration involves soft gluons
with energies of order a nominal scale (NS), collinear gluons with
energies of order the NS, and  collinear gluons with energy of the large
scale (LS). The LS is much larger than the NS, but much smaller than the
NS of the next larger (inner) level. We apply the soft and collinear
approximations and the decoupling relations at each stage. We also make
use of relationships between collinear eikonal lines to combine
contributions within each stage and to combine contributions from successive
stages. We refer the reader to Ref.~\cite{Bodwin:2009cb} for details.
The result is that the soft and collinear singular contributions factor,
and we arrive at the form that is shown in Fig.~\ref{fig:factored-form}.
In this figure, $\tilde S$, $\tilde{J}^\pm$ denote the singular parts of 
$S$ and $J^\pm$.

It can be seen that the soft eikonal lines that attach to a quark and an
antiquark in a given meson cancel. This can be established by making use
of order-by-order algebraic relations or, more simply, by noticing that
the corresponding path-ordered exponentials run in opposite directions
and end on space-time points that are separated by $k_\mu/Q\to 0$. The
cancelling contributions have the same color factor by virtue of the
color-singlet nature of the meson.

There is also a cancellation of the parts of the quark and antiquark
collinear eikonal lines for which the energies of the collinear gluons
are much less than $Q$. This cancellation implies that the couplings of
the low-energy $C^\pm$ gluons to subdiagrams outside $J^\pm$ do not
contribute in the end, but this becomes apparent only when one has
carried out the re-organization that we have described.

Now we extend the ranges of integration in $\tilde{J}^\pm$ up to an
ultraviolet cutoff $\mu_F\sim Q$, which acts as a factorization scale.
We also  re-define $\tilde{H}$ to be the complete amplitude divided by
$\tilde{J}^+$ and $\tilde{J}^-$. Then, $\tilde{H}$ is free of soft and
collinear singularities and depends only on the scale $Q$. Hence, we
have arrived at the standard factorized form:
\begin{equation}
A=\tilde{J}^-\otimes \tilde{H} \otimes \tilde{J}^+,
\end{equation}
where $\tilde{H}$ contains only virtualities of order $Q^2$ and
$\tilde{J}^+$ and $\tilde{J}^-$ contain all of the collinear
contributions with virtualities of order or less than $\Lambda_{\rm
QCD}^2$.

\begin{theacknowledgments}
The work of G.T.B.\ and X.G.T.\ was supported by the U.S. Department of
Energy, Division of High Energy Physics, under Contract
No.~DE-AC02-06CH11357. The research of X.G.T.\ was also supported by
NSERC. 
%The work of J.L.\ was supported
%by the Korea Ministry of Education, Science, and Technology through the
%National Research Foundation under Contract No.~2009-0086383.
\end{theacknowledgments}

%%
%% End of file `template-8d.tex'.
\end{document}